\documentclass[12pt] {article}
\usepackage{latexsym}{\rm }
\usepackage{graphics}
\usepackage{graphpap}
\newtheorem{lemma}{Lemma}
\newcommand{\beq}{\begin{equation}}
\newcommand{\feq}[1]{\label{#1} \end{equation}}
\newcommand{\beqr}{\begin{eqnarray}}
\newcommand{\feqr}{\end{eqnarray}}
\def\non{\nonumber}
\def\noi{\noindent}
\def\slasha#1{\setbox0=\hbox{$#1$}#1\hskip-\wd0\hbox to\wd0{\hss\sl/\/\hss}}

\def\slashb#1{\setbox0=\hbox{$#1$}#1\hskip-\wd0\dimen0=5pt\advance
       \dimen0 by-\ht0\advance\dimen0 by\dp0\lower0.5\dimen0\hbox
         to\wd0{\hss\sl/\/\hss}}

\newcommand{\rf}[1]{(\ref{#1})}

\def\cqg#1#2#3{Class. Quantum Grav. {\bf{#1}} (#2) #3}

\renewcommand{\thefootnote}{\fnsymbol{footnote}}

\begin{document}

\begin{center}


{\Large \bf The noncommutative harmonic oscillator in more than one dimensions}\\ [4mm]

\large{Agapitos Hatzinikitas}\footnote [2] 
{Email: ahatzini@tem.uoc.gr} \\ [5mm]

{\small University of Crete, \\
Department of Applied Mathematics, \\
L. Knosou-Ambelokipi, 71409 Iraklio Crete,\\ 
Greece}\\ [5mm]

\large{and} \\ [5mm]

\large{Ioannis Smyrnakis} \footnote [3] 
{Email: smyrnaki@tem.uoc.gr}\\ [5mm]

{\small University of Crete, \\
Department of Applied Mathematics, \\
L. Knosou-Ambelokipi, 71409 Iraklio Crete,\\ 
Greece}
\vspace{5mm}

\end{center}

\begin{abstract}

The noncommutative harmonic oscillator in arbitrary dimension is examined. It is shown that the $\star$-genvalue 
problem can be decomposed into separate harmonic oscillator equations for each dimension. The noncommutative plane
is investigated in greater detail. The constraints for rotationally symmetric solutions  and the 
corresponding two-dimensional harmonic oscillator are solved. The angular momentum operator is derived and its
$\star$-genvalue problem is shown to be equivalent to the usual eigenvalue problem. The $\star$-genvalues for the angular
momentum are found to depend on the energy difference of the oscillations in each dimension. 
Furthermore two examples of assymetric 
noncommutative harmonic oscillator are analysed. The first is the noncommutative two-dimensional Landau problem and the 
second is the three-dimensional harmonic oscillator with symmetrically noncommuting coordinates and momenta.

\end{abstract}
\newpage

\section{Introduction}
\renewcommand{\thefootnote}{\arabic{footnote}}
\setcounter{footnote}{0}

Noncommutative spaces, within the formulation of the deformation quantization through Moyal product, have been 
receiving increasing attention in field theories on $R^{d}$, string theories, in which they arise naturally
as D-brane configuration spaces, and M-theory.

There has been 
considerable progress towards the direction of understanding the interplay between the UV and IR divergencies 
steming from the underlying noncommutativity of scalar field theories on flat space-time. Some authors 
partially motivated 
by the suggestion of \cite{seiberg} on the possible stringy origin of the intriguing UV/IR mixing, have explored 
the relationship between noncommutative field theory and string theory \cite{andreev}.

In string theory it was realized recently that the introduction of a constant $B$-field gives rise to noncommutative 
string position operators. The multiloop amplitudes were computed for this noncommutative bosonic string in 
\cite{chu}.

The appearance of noncommutative spaces triggered the reformulation of Quantum Mechanics \cite{zachos}. 
In this set up 
the one dimensional Schr\"{o}dinger equation was replaced by the $\star$-genvalue equation,
while the wave functions became 
Wigner functions. The ordinary product was replaced by the pivotal associative noncommutative 
$\star$-product. 

In the present work we consider the $\star$-genvalue problem for the n-dimensional noncommutative harmonic 
oscillator.  It is shown that again the $\star$-genvalue problem is equivalent to the  Schr\"{o}dinger  problem 
in an appropriate representation.  The energy eigenvalues and eigenfunctions are determined as functions 
of the noncommutativity parameters.  

The case of the two dimensional harmonic oscillator is examined thoroughly.  The angular momentum operator 
is derived in the rotationally symmetric case.  It is shown that the $\star$-genvalues for this operator 
contain, apart from the usual angular momentum, a term that depends on the energy difference of the oscillations
in the two dimensions.   This is to be interpreted as the angular momentum of the system.

In the asymmetric case critical values of the parameters arise.  For these values the energy spectrum becomes 
infinitely degenerate at every level.  This problem can be identified with the noncommutative Landau 
problem.  

The paper is organized as follows:

In section 2 we summarize known results namely the definition of the Moyal 
product of functions in phase-space through the Weyl ordering 
prescription and the equivalence of the stationary $\star$-genvalue problem to the corresponding Schr\"{o}dinger problem.

Section 3 is dedicated to the study of the two-dimensional harmonic oscillator for nontrivial commutation 
relations. The $\star$-genvalues and functions are determined for the corresponding 
Hamiltonian by solving the imaginary and real part equations. 

In section 4 the $\star$-genvalue problem for the Hamiltonian and the angular momentum in two-dimensions  
is investigated. It is shown, that the Schr\"{o}dinger problem in suitably transformed phase-space variables  
is equivalent to the $\star$-genvalue problem. Furthermore the $\star$-genvalue problem for 
the angular momentum enables one to calculate the eigenvalues of the associated operator.

In section 5 the existence of rotationally symmetric 
$\star$-genfunctions impose constraints on the possible commutation relations for the n-dimensional harmonic 
oscillator. In two-dimensions these constraints are solved explicitly and the $\star$-genvalue problem for the 
Hamiltonian is solved. The angular momentum operator, which is the generator of rotations, is constructed and 
its  $\star$-genvalues are computed.

Finally section 6 copes with the most general commutation relations one can possibly encounter. 
We show that it is possible, through orthogonal transformations, to bring the matrix realizing the commutation 
relations 
into a symplectic form. In this way we can give explicit results for the energy levels, eigenfunctions and 
the $\star$-commutation relation of creation and annihilation operators for the n-dimensional harmonic oscillator. 
As a non-rotationally symmetric application 
we consider the noncommutative Landau problem and find a critical point for the magnetic field 
at which the energy levels are infinitely degenerate \cite{nair}. As a further application the three-dimensional noncommutative
harmonic oscillator is considered.         


\section{Overview of the Wigner functions and the $\star$-genvalue problem}

Let us start with a classical Hamiltonian $H(q,p)=p^2/2m+V(q)$ in one dimension. Upon quantization the canonical 
variables $q, p$ become operators $\hat{q} ,\hat{p} $ satisfying the canonical commutation relations 
$[\hat{q},\hat{p}]=i\hbar$.  Consider now monomials of the form $q^m p^n$ with m, n positive integers. 
To define the corresponding operator product it is possible to use the Weyl ordering prescription \cite{castel}: 
\beqr
\hat{W}(q^m p^n) &=& \frac{1}{2^n}\sum_{k=0}^n\left( \begin{array}{c} n \\ k  \end{array} \right)
\hat{p}^{n-k}\hat{q}^m\hat{p}^k \non \\
&=& \left[ e^{-\frac{i\hbar}{2}\frac{\partial^2}{\partial q \partial p}} q^m p^n \right]_{q\rightarrow \hat{q},
p\rightarrow \hat{p}}
\label{weylor}
\feqr
\noi according to which the $\hat{W}(q^m p^n)$ operator is symmetrized in $\hat{q}$ and $\hat{p}$ 
by use of Heisenberg's commutation relation. This regularization scheme can be extended to act on arbitrary power 
series functions $f(q,p)$ through linearity. Thus, 
Weyl ordering is an invertible map from the space of functions on the phase-space to the space of 
quantum operators.  We can use now Weyl ordering to define a new product between functions on the phase-space 
 \cite{zachos}, \cite{castel}:
\beqr
f(q,p)\star g(q,p) &=& W^{-1}(W(f)W(g)) \non \\
&=& f(q,p)e^{i\frac{\hbar}{2} \left(\overleftarrow{\partial}_{q}
\overrightarrow{\partial}_{p}
-\overleftarrow{\partial}_{p}\overrightarrow{\partial}_{q} \right)} g(q,p) \non \\
&=& f(q+i\frac{\hbar}{2} \overrightarrow{\partial}_{p} , p-i\frac{\hbar}{2} \overrightarrow{\partial}_{q})
g(q,p).
\label{moyal}
\feqr

\noi This is the celebrated Moyal product which enjoys the properties of noncommutativity, associativity and 
uniqueness.

\par  
In phase-space one can define Wigner quasi-distribution functions to calculate matrix elements of observables.  
The time-independent Wigner function corresponding to a pair of eigenstates $|\psi_{n}>$, $|\psi_{m}>$ 
of the Schr\"{o}dinger problem,
$\hat{H}|\psi_{n}>=E_{n}|\psi_{n}>$, is represented in two phase-space dimensions by:
\beqr
f_{mn}(q,p)= \frac{1}{2\pi} \int dy e^{-iyp} <q-\frac{\hbar}{2}y|\psi_{n}>
<\psi_{m}|q+\frac{\hbar}{2}y>
=f^{\ast}_{nm}(q,p)
\label{eq1}
\feqr
\noi where $\ast$ stands for complex conjugation. In this case, 
one can show by employing the classical Hamiltonian and the definition of the star product \rf{moyal} 
that Wigner functions obey the $\star$-genvalue equations:
\beqr
H(q,p) \star f_{mn}(q,p)=E_{n}f_{mn}(q,p) \quad ; \quad f_{mn}(q,p) \star H(q,p) =E_{m}f_{mn}(q,p).
\label{eq3}
\feqr
\noi In \rf{eq1} for complete sets $|\psi_{n}>$ one can also derive that:
\beqr
\sum_{m,n} f_{mn}(q,p)f^{\ast}_{mn}(q', p')= \frac{1}{2\pi \hbar} \delta(q-q')\delta(p-p')
\label{eq4}
\feqr     
\noi which enables the construction of arbitrary phase-space functions in terms of $f_{mn}(q,p)$.
\par For the diagonal case ($m=n$ in \rf{eq1}) one can prove that the Wigner functions satisfy the diagonal 
$\star$-genvalue equation:
\beqr
H(q,p) \star f_{n}(q,p)=f_{n}(q,p) \star H(q,p) =E_{n}f_{n}(q,p) 
\label{dstarg}
\feqr
\noi Furthermore, the real solutions of \rf{dstarg} 
are required to be of the Wigner type for the wavefunctions of $\hat{H}|\psi_{n}>=E_{n}|\psi_{n}>$.  So, instead of 
solving the Schr\"{o}dinger equation one can try to solve the $\star$-genvalue equation to determine directly the 
diagonal Wigner functions and the corresponding energy spectrum. The orthogonality relation satisfied by those 
functions takes the form: 
\beqr
f_{m}\star f_{n}=\frac{1}{2\pi \hbar}\delta_{m,n}f_{m}.
\label{eq5}
\feqr




\section{The two-dimensional harmonic oscillator}

In this case we have four coordinates $(q_i,p_j)$, with $i,j=1,2$, in phase space.  The free Hamiltonian for this 
model is: 
\beqr
H(q_i,p_j)=\frac{1}{2}\sum_{i=1}^{2} \left(q_i^2+p_i^2 \right)
\label{hamilt}
\feqr
\noi where, without loss of generality, parameters have been absorbed in the phase-space variables. 
Quantum Mechanically, the position and momentum operators
satisfy the Heisenberg commutation relations. Here we will extend this realization to include nontrivial 
commutation relations for $q_1,q_2$ and $p_1,p_2$: 
\beqr 
[q_i,p_j]=i\hbar \delta_{ij} \quad ; \quad [q_1,q_2]=i\theta \quad ; \quad [p_1,p_2]=-i\theta
\label{commut}
\feqr
\noi where $\theta $ is a real constant.  The related star product takes the form: 
\beqr
\star= \exp \left[ \frac{i}{2} \left( \begin{array}{cccc} \overleftarrow{\partial}_{q_1}, &
\overleftarrow{\partial}_{p_1}, & \overleftarrow{\partial}_{q_2}, &
\overleftarrow{\partial}_{p_2} \end{array} \right) 
\left( \begin{array}{cccc} 0 & \hbar & \theta & 0 \\ 
-\hbar & 0 & 0 & -\theta \\
-\theta & 0 & 0 & \hbar \\
0 & \theta & -\hbar &0 \end{array} \right) 
\left( \begin{array}{cccc} \overrightarrow{\partial}_{q_1} \\
\overrightarrow{\partial}_{p_1} \\ \overrightarrow{\partial}_{q_2} \\
\overrightarrow{\partial}_{p_2} \end{array} \right)
\right].
\label{star2}
\feqr
\noi The resulting $\star$-genvalue equation is: 
\beqr
&\Bigg[& \!\!\!\!\!\!   \left(q_1+\frac{i\hbar}{2}\partial_{p_1}+\frac{i\theta}{2}\partial_{q_2}\right)^2 
\! + \!
\left(p_1-\frac{i\hbar }{2}\partial_{q_1}-\frac{i\theta}{2}\partial_{p_2}\right)^2 \non \\
&+& \!\!\!\!\!\!  \left(q_2+\frac{i\hbar }{2}\partial_{p_2}-\frac{i\theta}{2}\partial_{q_1}\right)^2 
\! + \!
\left(p_2-\frac{i\hbar }{2}\partial_{q_2}+\frac{i\theta}{2}\partial_{p_1}\right)^2 \Bigg] \!\!
\!\! f(q_1,p_1,q_2,p_2) \! = \! 2Ef(q_1,p_1,q_2,p_2).
\label{2dsgen}
\feqr
\noi Equation \rf{2dsgen} splits into an equation for the imaginary part, 
\beqr
\Bigg[ \hbar(p_1\partial_{q_1}-q_1\partial_{p_1})+\hbar(p_2\partial_{q_2}-q_2\partial_{p_2})+
\theta \left(q_2\partial_{q_1}-q_1\partial_{q_2}\right)+
\theta \left(p_1\partial_{p_2}-p_2\partial_{p_1}\right) \Bigg]f=0
\label{imeq}
\feqr
\noi and an equation for the real part:
\beqr
\left[ \left(q_1^2+p_1^2+q_2^2+p_2^2 \right)-\frac{(\hbar^2+\theta^2)}{4}
\left(\partial_{q_1}^2+\partial_{p_1}^2+
\partial_{q_2}^2+\partial_{p_2}^2 \right) \right]f=2Ef.
\label{requ}
\feqr
\noi Equation \rf{imeq} admits a solution of the form $f(z)$ with $z=2(q_1^2+p_1^2+q_2^2+p_2^2)=4H$.  
The real part equation \rf{requ} transforms to the ordinary differential equation:
\beqr
\left[ z\partial_z^{2}+2\partial_z+\frac{1}{(\hbar^2+\theta^2)} \left(E-\frac{z}{4}\right) \right] f(z)=0.
\label{requ1}
\feqr
\noi The problem with this reduction is the inconsistency underlying the number of degrees of freedom. We started 
with the two dimensional harmonic oscillator and we ended up with one oscillation equation.  So we 
need to search for a set of transformations that will preserve the number of degrees of freedom.  The 
appropriate transformations are:
\beqr
\begin{array}{cc}
\bar{q_1}=q_1 & \bar{p_1}=\frac{1}{\sqrt{\hbar^2+\theta^2}}(\hbar p_1+\theta q_2) \\
\bar{q_2}=\frac{1}{\sqrt{\hbar^2+\theta^2}}(\hbar q_2-\theta p_1) & \bar{p_2}=p_2.
\end{array}
\label{transfs}
\feqr
\noi The imaginary part equation is then written as:
\beqr
\Bigg[ \left(\bar{p}_1\partial_{\bar{q}_1}-\bar{q}_1\partial_{\bar{p}_1} \right)+
\left( \bar{p}_2\partial_{\bar{q}_2}-\bar{q}_2\partial_{\bar{p}_2} \right) \Bigg]f=0.
\label{imeq1}
\feqr
\noi This implies that $f$ is a function of $z_1=2(\bar{q}_1^2+\bar{p}_1^2)$ and $z_2=2(\bar{q}_2^2+\bar{p}_2^2)$.  
The real part equation transforms under the new variables into:
\beqr
\left[ z_{1}\partial_{z_1}^{2}+\partial_{z_1}+z_{2}\partial_{z_2}^{2}+\partial_{z_2}+
\frac{1}{(\hbar^2+\theta^2)}\left(E-\frac{(z_1+z_2)}{4} \right) \right]f(z_1,z_2)=0.
\label{requ2}
\feqr
\noi The diagonal $\star$-genfunctions $f_{nm}(z_1,z_2)$ associated with this equation are
determined through products of Laguerre polynomials: 
\beqr
f_{nm}(\tilde{z}_1,\tilde{z}_2)=e^{-\frac{1}{2}(\tilde{z}_1 +\tilde{z}_2)}
L_n(\tilde{z}_1)L_m(\tilde{z}_2)
\label{eigfuncts}
\feqr
\noi where:
\beqr
L_m(\tilde{z}_i)=\frac{1}{m!}e^{\tilde{z}_i}\frac{d^m}{d^{m} \tilde{z}_i} 
\left(e^{-\tilde{z}_i} \tilde{z}_{i}^{m} \right)
\label{lague}
\feqr
\noi and $L_0(\tilde{z_i})=1$, $L_1(\tilde{z_i})=1-\tilde{z_i}$, 
$L_2(\tilde{z_i})=1-2\tilde{z_i}+\frac{\tilde{z_i}^2}{2}, \cdots$ with $\tilde{z_i}=z_i/\sqrt{\hbar^2+\theta^2}$. 
The energies corresponding to these $\star$-genfunctions are: 
\beqr
E_{nm}=\sqrt{\hbar^2+\theta^2} \left(n+m+1 \right).
\label{eigvals}
\feqr
The annihilation and creation operators are given in terms of the transformed variables as: 
\beqr
a_i=\frac{1}{\sqrt{2}}(\bar{q}_i +i\bar{p}_i)\qquad 
a_i^{\dagger }=\frac{1}{\sqrt{2}}(\bar{q}_i -i\bar{p}_i).
\label{annih1}
\feqr
\noi It is possible to express these operators in terms of the original variables by reversing our transformations.
They satisfy the following modified commutation relations
\beqr
a_i \star a_j^{\dagger}-a_j^{\dagger} \star a_i = \delta_{ij} \sqrt{\hbar^2+\theta^2}
\label{com1}
\feqr
\noi and $a_i\star f_{00}(\bar{q}_i,\bar{p}_i)=0$. So they generate the $\star$-Fock space of states as follows:
\beqr
f_{nm} \propto a_{1}^{\dagger n} a_{2}^{\dagger m} \star f_{00} \star a_{2}^{m} a_{1}^{n}.
\label{fock}
\feqr
The Hamiltonian takes the following 
form in terms of annihilation-creation operators:
\beqr
H=\sum_{i=1}^2\left( a^{\dagger}_i\star a_i+\frac{\sqrt{\hbar^2+\theta^2}}{2}\right). 
\label{hamilt3}
\feqr




\section{Quantum Mechanics and the $\star$-genvalue problem in 2D}
In the one dimensional case the $\star$-genvalue equation is equivalent to the Schr\"{o}dinger  equation, 
as was shown in \cite{zachos}.  However, in the two dimensional case one has to search for a suitable 
representation 
of the commutation relations before writing down an equation.  This problem can be overcome for the 
harmonic oscillator of section 3 by using the transformations \rf{transfs}, to transform the 
commutation relations to two sets of Heisenberg commutation relations with 
$\hbar $ replaced by $\sqrt{\hbar^2+\theta^2}$. The 
Schr\"{o}dinger equation with respect to the new variables becomes:
\beqr
\frac{1}{2}\sum_{i=1}^{2} \left(\hat{\bar{q}}_i^2+\hat{\bar{p}}_i^2 \right)\psi(\bar{q}_i,\bar{p}_i)
=E\psi(\bar{q}_i,\bar{p}_i)
\label{Schr}
\feqr
\noi and its eigenfunctions are given by:
\beqr
\psi_{nm}(\bar{q}_i,\bar{p}_i)=\psi_n(\bar{q}_1,\bar{p}_1) \psi_m(\bar{q}_2,\bar{p}_2)
\label{eigfcts}
\feqr
\noi where $\psi_n$ are the usual eigenfunctions of the one dimensional harmonic oscillator.  
This equation splits into two one-dimensional harmonic oscillator equations. 
Making use of the one-dimensional equivalence 
of the  Schr\"{o}dinger  problem to the  $\star$-genvalue  problem we obtain that equation \rf{Schr}  
(with $\hbar $ replaced by $\sqrt{\hbar^2+\theta^2}$) is equivalent to:
\beqr
H(\bar{q}_i,\bar{p}_i)\bar{\star} f(\bar{q}_i,\bar{p}_i)=f(\bar{q}_i,\bar{p}_i)\bar{\star} H(\bar{q}_i,\bar{p}_i)=
E f(\bar{q}_i,\bar{p}_i).
\label{2dstar}
\feqr
The $\bar{\star} $-product, which is the transformed version of the $\star $-product defined in \rf{star2}, is: 
\beqr
\bar{\star}=\exp \left[ \frac{i}{2} \sqrt{\hbar^2+\theta^2}\sum_{i=1}^2(\overleftarrow{\partial }_{\bar{q}_i}
\overrightarrow{\partial }_{\bar{p}_i}-\overleftarrow{\partial }_{\bar{p}_i}
\overrightarrow{\partial }_{\bar{q}_i})\right]. 
\label{bstar}
\feqr
\noi Going back to the original variables the Schr\"{o}dinger  problem  is translated into the 
$\star$-genvalue  problem of section 3.  
The diagonal Wigner functions are given in terms of the wave functions by:
\beqr
f(\bar{q}_i , \bar{p}_j)&=&  f(\bar{q}_1 , \bar{p}_1) f(\bar{q}_2 , \bar{p}_2) \non \\
&=& \frac{1}{\left(2\pi \right)^2}\int \int dy_1 \,\, dy_2 \,\, e^{-iy_1 \bar{p}_1} \,\,
e^{-iy_2 \bar{p}_2} \,\, \prod_{i=1}^{2} 
\Psi^{\ast}_{i}(\bar{q}_i - \frac{1}{2} \sqrt{\hbar^2+\theta^2}y_i)
\Psi_{i}(\bar{q}_i + \frac{1}{2} \sqrt{\hbar^2+\theta^2}y_i) \non \\
&=& \frac{1}{\left(2\pi \right)^2}\int \int dy_1 \,\, dy_2 \,\, e^{-iy_1 \bar{p}_1} \,\,
e^{-iy_2 \bar{p}_2} \,\, \prod_{i=1}^{2} \Psi^{\ast}_{i} \Psi_{i}.
\label{wf2d}
\feqr
\noi Expressing the Wigner functions in the original variables one has: 
$f_0(q_i,p_i)\equiv f(\bar{q}_i,\bar{p}_i)$.

\par  Next let us investigate what happens with the angular momentum.  The operator that commutes with the 
Hamiltonian and generates rotations in the original variables is:
\beqr
L=\frac{\hbar}{\hbar^2+\theta^2}\left[ \hbar(q_1p_2-q_2p_1)+\frac{\theta}{2}(p_1^2-q_1^2+p_2^2-q_2^2)\right]. 
\label{angm}
\feqr
\noi Transforming this to the new variables gives:
\beqr
\bar{L}(\bar{q}_i,\bar{p}_j) = \frac{\hbar}{\hbar^2+\theta^2} \left[ \hbar (\bar{q}_1 \bar{p}_2 -\bar{q}_2 \bar{p}_1) 
+\frac{\theta}{2} \left(\bar{q}^{2}_{2}+\bar{p}^{2}_{2}-\bar{q}^{2}_{1}-\bar{p}^{2}_{1} \right)\right].
\label{angmn}
\feqr
\noi The angular momentum $\star$-genvalue equation becomes:
\beqr
&L& \!\!\!\! (q_i , p_i) \star f_0(q_i , p_i)=
\bar{L}(\bar{q}_i , \bar{p}_i) \bar{\star} f(\bar{q}_i , \bar{p}_i) \non \\
\!\!\!\! &=& \!\!\!\! \frac{1}{\left(2\pi \right)^2}
\frac{\hbar}{(\hbar^2+\theta^2)} \Bigg[ \hbar \left[ \bar{q}_1 \left(\bar{p}_2- \frac{i}{2}\sqrt{\hbar^2+\theta^2} 
\overrightarrow{\partial}_{\bar{q}_2} \right) 
- \bar{q}_2 \left(\bar{p}_1- \frac{i}{2}\sqrt{\hbar^2+\theta^2}\overrightarrow{\partial}_{\bar{q}_1} \right)\right] \non \\
\!\!\!\! &+& \!\!\!\! \frac{\theta}{2} \Bigg[\bar{q}^{2}_{2}+ 
\left(\bar{p}_2- \frac{i}{2}\sqrt{\hbar^2+\theta^2}\overrightarrow{\partial}_{\bar{q}_2} \right)^2 -
\bar{q}^{2}_{1}- \left( \bar{p}_1- \frac{i}{2}\sqrt{\hbar^2+\theta^2}\overrightarrow{\partial}_{\bar{q}_1}\right)^2
\Bigg] \Bigg] \non \\
\!\!\!\! &\times& \!\!\!\! \int dy_1 \,\, dy_2 \,\, 
e^{-iy_1 \left(\bar{p}_1 + \frac{i}{2} \sqrt{\hbar^2+\theta^2} \overleftarrow{\partial}_{\bar{q}_1} \right)} \,\,
e^{-iy_2 \left( \bar{p}_2 + \frac{i}{2} \sqrt{\hbar^2+\theta^2} \overleftarrow{\partial}_{\bar{q}_2}\right)} 
\,\, \prod_{i=1}^{2} \Psi^{\ast}_{i} \Psi_{i} \non \\
\!\!\!\! &=& \!\!\!\! \frac{1}{\left(2\pi \right)^2} \frac{\hbar}{(\hbar^2+\theta^2)}
\int \int dy_1 \,\, dy_2 \,\, e^{-iy_1 \bar{p}_1} \,\, e^{-iy_2 \bar{p}_2} \non \\
\!\!\!\! &\times& \!\!\!\!   \Bigg[ \hbar \Bigg[
\left(\bar{q}_1 +  \frac{1}{2}\sqrt{\hbar^2+\theta^2}y_1  \right) 
\left(-i\partial_{y_2}- \frac{i}{2}\sqrt{\hbar^2+\theta^2} 
\overrightarrow{\partial}_{\bar{q}_2} \right) \non \\
\!\!\!\! &-& \!\!\!\! \left(\bar{q}_2 +  \frac{1}{2}\sqrt{\hbar^2 +\theta^2}y_2  \right) 
\left(-i\partial_{y_1}- \frac{i}{2}\sqrt{\hbar^2+\theta^2} 
\overrightarrow{\partial}_{\bar{q}_1} \right)\Bigg] \non \\
\!\!\!\! &+& \!\!\!\! \frac{\theta}{2} \Bigg[ \left(\bar{q}_2 +  \frac{1}{2}\sqrt{\hbar^2+\theta^2}y_2  \right)^2 
+ \left(-i\partial_{y_2}- \frac{i}{2}\sqrt{\hbar^2+\theta^2} 
\overrightarrow{\partial}_{\bar{q}_2} \right)^2  \non \\
\!\!\!\! &-& \!\!\!\! \left(\bar{q}_1 +  \frac{1}{2}\sqrt{\hbar^2+\theta^2}y_1  \right)^2 
- \left(-i\partial_{y_1}- \frac{i}{2}\sqrt{\hbar^2+\theta^2} 
\overrightarrow{\partial}_{\bar{q}_1} \right)^2 \Bigg] \Bigg] \,\, \prod_{i=1}^{2} \Psi^{\ast}_{i} \Psi_{i} \non \\
\!\!\!\! &=& \!\!\!\! \frac{\hbar}{\left(\hbar^2+ \theta^2 \right)} \left[\hbar \sqrt{\hbar^2 + \theta^2}m_{z}+\theta 
\left(E_2 - E_1\right) \right] f(\bar{q}_i , \bar{p}_j).
\label{equiss}
\feqr
\noi Here we have used $p_ie^{-iy_ip_i}=i\partial_{y_i}e^{-iy_i p_i}$ and partial integration has been 
performed. Note that: $$\left( \partial_{y_i}+ \frac{1}{2}\sqrt{\hbar^2+\theta^2} 
\overrightarrow{\partial}_{\bar{q}_i} \right)
\Psi^{\ast} \left(\bar{q}_i-\frac{\sqrt{\hbar^2+\theta^2}}{2}y_i \right)=0.$$  
The discrete values $m_{z}$ are the eigenvalues of the angular momentum for the commutative two dimensional harmonic oscillator:
\beqr
\hat{L} \Psi(\bar{q}_i , \bar{p}_j) =
\left(\hat{\bar{q}}_1 \hat{\bar{p}}_2 -\hat{\bar{q}}_2 \hat{\bar{p}}_1 \right) 
\Psi(\bar{q}_i , \bar{p}_i) = \sqrt{\hbar^2 + \theta^2} m_z \Psi(\bar{q}_i , \bar{p}_i).  
\label{mang}
\feqr
What we 
have essentially proven is that the solution of the $\star$-genvalue problem for the angular momentum is equivalent 
to the solution of the eigenvalue problem for the angular momentum operator. In the 
noncommutative case we pick an extra new term that depends on the energy difference of the two oscillations 
and the noncommutativity parameter $\theta$.



\section{Rotationally symmetric case}
\par Up to this point we have considered only the case where $q_i$ commute with $p_j$, for $i\ne j$ and the 
$q_i$, $p_j$ behave symmetrically, that is $[q_1,q_2]=-[p_1,p_2]$.  Lets consider the more general situation 
where the commutation relations are governed by a general antisymmetric matrix $M$.  
In this case the $\star $-product reads:
\beqr
\star=\exp \frac{i}{2} \left[ \overleftarrow{\partial }_I^T M_{IJ}\overrightarrow{\partial}_J\right] 
\label{star3}
\feqr
\noi where $\partial_I^T=(\partial_{q_I},\partial_{p_I})$, and $M_{IJ}=-M_{JI}^T$ are $2\times 2$ matrices.  
Here the summation convention is assumed.  
The imaginary part of the $\star$-genvalue equation now becomes: 
\beqr
X_I^TM_{IJ}\partial_Jf=0
\label{imeq3}
\feqr
\noi where $X_I^T=(q_I,p_I)$.  If $f\equiv f(X_I^TX_I)$ then this equation is satisfied provided that $M$ is
antisymmetric.  Interestingly enough, 
if one starts with the $\star$-genvalue problem for this $f$, ignoring the commutation relations,  
equation \rf{imeq3} would require that the matrix $M$ be antisymmetric.  
\par The real part equation turns into the form: 
\beqr
\left[ X_I^TX_I-\frac{1}{4}(M_{IK_1}\partial_{K_1})^T(M_{IK_2}\partial_{K_2}) \right] f=2Ef.
\label{requ4}
\feqr
\noi Again demanding $f=f(X_I^TX_I)$ we are led to: 
\beqr
 X_I^TX_If(X_I^TX_I) \! - \! \frac{1}{2}tr(M_{IK}M_{IK})f'(X_I^TX_I) \! - \!
 X_{K_1}^TM_{IK_1}^TM_{IK_2}X_{K_2}f''(X_I^TX_I) \! = \! 2Ef(X_I^TX_I).
 \label{requ5}
 \feqr
 \noi If this equation is to have rotationally symmetric solutions we need: 
 \beqr
  M_{IK_1}^TM_{IK_2}=\alpha \delta_{K_1K_2}I
  \label{cond}
  \feqr
  \noi with $I$ the identity matrix.
  
  \par Condition \rf{cond} can be solved explicitly in the case of four phase-space dimensions.  The most general 
  matrix $M$ that admits rotationally symmetric $\star$-genfunctions is given by:
  \beqr
  M_{11}=\pm M_{22}=\left( \begin{array}{cc} 0 & \hbar \\ -\hbar & 0 \end{array} \right) \quad ; \quad
  M_{12}=-M_{21}=\left( \begin{array}{cc} \theta & \phi \\ \pm \phi & \mp \theta \end{array} \right) 
  \label{matr}
  \feqr
  \noi where the two signs correspond to the two possible solutions that exist. We are going to 
  examine more closely the first case, since the second can be treated on equal footing.  
  In the present case the transformations that lead to the Heisenberg commutation relations are: 
  $\bar{X}=R^T X$, where $R$ is given by:
  \beqr
  R=\left( \begin{array}{cccc}\frac{\sqrt{\hbar^2+\phi^2}}{\sqrt{\hbar^2+\theta^2+\phi^2}} & 0 & 0 &
  -\frac{\theta}{\sqrt{\hbar^2+\theta^2+\phi^2}} \\
  -\frac{\theta \phi}{\sqrt{\hbar^2+\theta^2+\phi^2}\sqrt{\hbar^2+\phi^2}} & 
  \frac{\hbar}{\sqrt{\hbar^2+\phi^2}} & 0 & -\frac{\phi}{\sqrt{\hbar^2+\theta^2+\phi^2}} \\
  0 & 0 & 1 & 0 \\
 \frac{\hbar \theta }{\sqrt{\hbar^2+\theta^2+\phi^2}\sqrt{\hbar^2+\phi^2}} & 
  \frac{\phi}{\sqrt{\hbar^2+\phi^2}} & 0 & \frac{\hbar}{\sqrt{\hbar^2+\theta^2+\phi^2}} \end{array} \right)
  \label{tranmat}
  \feqr
 \noi Certainly these transformations are not unique. They are defined up to symplectic rotations which 
 preserve both the Hamiltonian and the commutation relations in the transformed variables. 
 \par The angular momentum is found to be:
\beqr
L=\frac{\hbar}{\hbar^2+\theta^2 + \phi^2}\left[ \hbar(q_1p_2-q_2p_1)+\frac{\theta}{2}(p_1^2-q_1^2+p_2^2-q_2^2)
-\phi(p_1 q_1 + p_2 q_2)\right]
\label{angn}
\feqr
\noi and in the transformed variables is reexpressed as: 
\beqr
\bar{L}=\frac{\hbar}{\hbar^2+\theta^2 + \phi^2} \left[ \sqrt{\hbar^2+ \phi^2} (\bar{q}_1 \bar{p}_2-\bar{q}_2 \bar{p}_1)
+\frac{\theta}{2} ( \bar{q}_1^2+\bar{p}_1^2- \bar{q}_2^2-\bar{p}_2^2)\right].
 \label{angm2}
 \feqr

\par  Again the $\star$-genvalue problem for the Hamiltonian is the same as the Schr\"{o}dinger problem with 
$\star$-genvalues $E_{n_1, n_2}= \sqrt{\hbar^2+\theta^2 + \phi^2}(n_1+n_2+1)$. The same 
equivalence is
valid for the angular momentum $\star$-genvalue problem producing the $\star$-genvalues  

$$\frac{\hbar}{\left(\hbar^2+ \theta^2+\phi^2 \right)} \left[\sqrt{\hbar^2 + \phi^2} \sqrt{\hbar^2 + \theta^2 + \phi^2}
 m_{z}+\theta 
\left(E_1 - E_2\right) \right]. $$ 

In the limit $\theta ,\phi \rightarrow 0$ we can recover from these expressions 
the usual ones.  




\section{General Case}
Let us assume now that the matrix $M$ used to define the commutation relations is a general antisymmetric matrix.  
The following lemma holds \cite{dusa}:
\begin{lemma}
Let $(V,\omega )$ be a symplectic vector space and $g:V\times V\rightarrow R$ be an inner product.  Then 
there exists a basis $u_1,\cdots ,u_n$, $v_1,\cdots ,v_n$ of $V$ which is both $g$-orthogonal and 
$\omega $-standard.  Moreover, this basis can be chosen such that $g(u_j,u_j)=g(v_j,v_j)$ for all j.  
\end{lemma}
This means that it is possible, by rescaling, to find an orthogonal transformation $R$ so that:  
\beqr 
R^TMR=J(M)
\label{mat}
\feqr
\noi where 
\beqr
J(M)_{IJ}=\alpha_I \delta_{IJ} \left( \begin{array}{cc} 0 & 1 \\ -1 & 0 \end{array}\right). 
\label{sympl}
\feqr
\par 
From \rf{star3} we see that if we make the transformation $\bar{X}=R^TX$ , then the matrix $M$ in the 
$\star$-product is replaced by $J(M)$ and the Hamiltonian remains invariant because the transformation 
is orthogonal.  So the $\star$-genvalue problem now becomes:
\beqr
\bar{H}\bar{\star }f=Ef
\label{starg2}
\feqr
where $\bar{\star }$ is
\beqr
\bar{\star}=\exp \left[ \frac{i}{2}  \overleftarrow{\bar{\partial} }_I^T J(M)_{IK}
\overrightarrow{\bar{\partial}}_K\right]. 
\label{star4}
\feqr
The imaginary part equation \rf{imeq3} becomes:
\beqr
\sum_{i} \alpha_i(\bar{q}_i\partial_{\bar{p}_i}-\bar{p}_i\partial_{\bar{q}_i})f(\bar{q}_i,\bar{p}_i)=0  
\label{imeq6}
\feqr
\noi where $\alpha_{i}=\alpha_{I}$. This equation is satisfied by $f\equiv f(z_i)$ where 
$z_i=2(\bar{q}_i^2+\bar{p}_i^2)$.
For this $f$, the real part equation takes the form:
\beqr
\sum_i \left[ z_i\partial^2_{z_i}+\partial_{z_i}-\frac{1}{\alpha_i^2 }(\frac{z_i}{4}-E_i) \right] f(z_i)=0
\label{req6}
\feqr
\noi where $E=\sum_{i} E_i$. 
This equation can be separated into a set of equations for each  $z_i$.  Solving these equations we get the 
eigenvalues $E_i=\alpha_i(n_i+1/2)$ and the eigenfunctions: 
\beqr
f_{n_i}(z_i)=e^{-\frac{1}{2\alpha_i}z_i}L_{n_i}(z_i/\alpha_i).
\label{eigfuncts1}
\feqr
The overall eigenfunctions are products of the $f_{n_i}(z_i)$.
The annihilation and creation operators again take the form \rf{annih1}.  They satisfy the following 
commutation relations:
\beqr
a_i \star a_j^{\dagger}-a_j^{\dagger} \star a_i = \alpha_i \delta_{ij} 
\label{com2}
\feqr
\noi ans the Hamiltonian becomes:
\beqr
H=\sum_i\left( a^{\dagger}_i\star a_i+\frac{\alpha_i}{2}\right). 
\label{hamilt5}
\feqr

Again, the $\star$-genvalue problem is equivalent to the Schr\"{o}dinger problem in the transformed variables. 
This means that it is also equivalent in the original variables, if one uses the representation for the original
variables that result from the usual representation of the transformed ones.  

\par As an example of a non-rotationally symmetric case let us consider the noncommutative Landau problem. 
Assume for convenience that $\hbar=1$.  Here 
the commutation relations are: 
\beqr 
[q_i,p_j]=i\delta_{ij} \quad ; \quad [q_1,q_2]=i\theta \quad ; \quad [p_1,p_2]=iB.
\label{commut5}
\feqr
The star operator is: 
\beqr
\star= \exp \left[ \frac{i}{2} \left( \begin{array}{cccc} \overleftarrow{\partial}_{q_1}, &
\overleftarrow{\partial}_{p_1}, & \overleftarrow{\partial}_{q_2}, &
\overleftarrow{\partial}_{p_2} \end{array} \right) 
\left( \begin{array}{cccc} 0 & 1 & \theta & 0 \\ 
-1 & 0 & 0 & B \\
-\theta & 0 & 0 & 1 \\
0 & -B & -1 &0 \end{array} \right) 
\left( \begin{array}{cccc} \overrightarrow{\partial}_{q_1} \\
\overrightarrow{\partial}_{p_1} \\ \overrightarrow{\partial}_{q_2} \\
\overrightarrow{\partial}_{p_2} \end{array} \right)
\right].
\label{star8}
\feqr
The $\alpha_I$ take the form: 
\beqr
\alpha_{\pm}=\frac{1}{2}\left( \sqrt{(\theta-B)^2+4}\pm (\theta +B)\right).
\label{freq}
\feqr
These $\alpha_{\pm }$ correspond to the frequencies of the Landau harmonic oscillator.  In this case there 
is no degeneracy in the $\alpha_{\pm }$ as opposed to the rotationally symmetric case.  
\par The transformation matrix R that corresponds to this case is

\beqr
R=\left(\begin{array}{cccc} -\frac{\alpha_-(1+ B\tilde{\alpha}_- )}{\sqrt{(1-B\theta )^2+
\alpha_-^2(1+B\tilde{\alpha}_- )^2}} & 0 & -\frac{\alpha_+(1-B\tilde{\alpha}_+ )}{\sqrt{(1-B\theta )^2+
\alpha_+^2(1-B\tilde{\alpha}_+ )^2}}  & 0 \\
0 & \frac{1}{\sqrt{1+\tilde{\alpha}_-^2}} & 0 &  \frac{1}{\sqrt{1+\tilde{\alpha}_+^2}}  \\
0 & -\frac{\tilde{\alpha }_-}{\sqrt{1+\tilde{\alpha}_-^2}} & 0 & \frac{\tilde{\alpha }_+}{\sqrt{1+\tilde{\alpha}_+^2}}\\
-\frac{1-B\theta }{\sqrt{(1-B\theta )^2+\alpha_-^2(1+B\tilde{\alpha}_- )^2}}  & 0 &
\frac{1-B\theta }{\sqrt{(1-B\theta )^2+\alpha_+^2(1-B\tilde{\alpha}_+ )^2}} & 0 
\end{array}\right)
\label{nonrot}
\feqr
\noi where we have assumed that $B\theta <1$ and 
\beqr
\tilde{\alpha}_{\pm}=\frac{1}{2}\left( \sqrt{(\theta-B)^2+4}\pm (\theta -B)\right).
\label{freq1}
\feqr
Note that if $B\theta =1$ then the fourth row in the transformation matrix becomes zero, so the transformation 
becomes degenerate, which is not permitted.  So there is a critical value for the magnetic field $B_0=1/\theta$.  
At $B_0$ the frequency $\alpha_-=0$, so there is an infinite degeneracy at the energy levels, corresponding to 
the excitations of the $\alpha_-$ oscillator.  
This problem was solved through a series of transformations in \cite{nair}, without resorting to the 
$\star$-genvalue formalism.  

\par As a further example consider the case of the noncommutative Landau problem with the additional commutation 
relations $[q_1,p_2]=i\phi_1$ and $[p_1,q_2]=i\phi_2$. In this case the frequencies become:
\beqr
\alpha_{\pm}=\frac{1}{2}\left( \sqrt{(B-\theta)^2+4+(\phi_1+\phi_2)^2}\pm \sqrt{(B+\theta)^2+(\phi_1-\phi_2)^2}\right)
\label{freq2}
\feqr
and a degeneracy is produced provided that: 
\beqr
B\theta-\phi_1\phi_2=1.
\label{cond2}
\feqr
Note that in this case the matrix M in the $\star $-product becomes degenerate again.  

\par As a final example consider the six phase space dimensional case. The commutation relations we consider 
are found to be:
\beqr
[q_i,p_j]=i\delta_{ij}\quad ;\quad [q_1,q_2]=i\theta_3\quad ; \quad [q_1,q_3]=-i\theta_2\quad ; \quad 
[q_2,q_3]=i\theta_1.
\label{comm9}
\feqr
\noi All other commutation relations are trivial.  Now the frequencies $\alpha_I$ are 
\beqr
\alpha_{1,2}=\sqrt {1+\theta_1^2+\theta_2^2+\theta_3^2}\quad , \quad \alpha_3=1.
\label{freq5}
\feqr
There is twofold degeneracy for the first frequency.  
The transformation matrix in this case is:
\beqr
R= \left( \begin{array}{cccccc} -\frac{\theta_1 \theta_3}{\alpha \beta} & \frac{\theta_2}{\beta} & 0 & 
\frac{\theta_1 \theta_3}{\alpha \beta \gamma} & 0 & -\frac{\theta_1}{\gamma} \\
-\frac{\theta_2}{\alpha \beta} & 0 & -\frac{\theta_1 \theta_3}{\beta \gamma} & 
-\frac{\theta_2 \gamma}{\alpha \beta} & \frac{\theta_1}{\gamma} & 0 \\
-\frac{\theta_2 \theta_3}{\alpha \beta} & -\frac{\theta_1}{\beta} & 0 & 
\frac{\theta_2 \theta_3}{\alpha \beta \gamma} & 0 & -\frac{\theta_2}{\gamma} \\
\frac{\theta_1}{\alpha \beta} & 0 & -\frac{\theta_2 \theta_3}{\beta \gamma} & 
\frac{\theta_1 \gamma}{\alpha \beta} & \frac{\theta_2}{\gamma} & 0 \\
\frac{\beta}{\alpha} & 0 & 0 & -\frac{\beta}{\alpha \gamma} & 0 & -\frac{\theta_3}{\gamma} \\
0 & 0 & \frac{\beta}{\gamma} & 0 & \frac{\theta_3}{\gamma} & 0  
\end{array} \right)
\label{tran6}
\feqr
\noi where 
\beqr
\alpha &=& \sqrt{1+\theta_1^2 +\theta_2^2 +\theta_3^2 } \non \\
\beta &=& \sqrt{\theta_1^2 + \theta_2^2} \non \\ 
\gamma &=& \sqrt{\theta_1^2 + \theta_2^2 + \theta_3^2}. \non
\label{defin2}
\feqr

    												





\bibliographystyle{plain}

\end{document}